\documentclass[12pt, preprint]{aastex}

\usepackage{graphicx}

\shortauthors{Kong et al.}

\usepackage{color}
\usepackage[normalem]{ulem}

\usepackage{multirow}

\begin{document}

\title{Observation of a metric type N solar radio burst}

\author{Xiangliang Kong\altaffilmark{1},
Yao Chen\altaffilmark{1},  Shiwei Feng\altaffilmark{1}, Guohui Du\altaffilmark{1}, Chuanyang Li\altaffilmark{1},
Artem Koval\altaffilmark{1,2}, V. Vasanth\altaffilmark{1}, Bing Wang\altaffilmark{1},
Fan Guo\altaffilmark{3}, and  Gang Li\altaffilmark{4}}

\altaffiltext{1}{Shandong Provincial Key Laboratory of Optical Astronomy and Solar-Terrestrial Environment,
and Institute of Space Sciences, Shandong University, Weihai, Shandong 264209, China; kongx@sdu.edu.cn}
\altaffiltext{2}{Institute of Radio Astronomy, National Academy of Sciences of Ukraine, Kharkiv 61002, Ukraine}
\altaffiltext{3}{Theoretical Division, Los Alamos National Laboratory, Los Alamos, NM 87545, USA}
\altaffiltext{4}{Department of Space Science and CSPAR, University of Alabama in Huntsville, Huntsville, AL 35899, USA}

\begin{abstract}
Type III and type-III-like radio bursts are produced by energetic electron beams guided along coronal magnetic fields.
As a variant of type III bursts, Type N bursts appear as the letter ``N" in the radio dynamic spectrum and
reveal a magnetic mirror effect in coronal loops.
Here, we report a well-observed N-shaped burst consisting of three successive branches
at metric wavelength with both fundamental and harmonic components
and a high brightness temperature ($>$10$^9$ K).
We verify the burst as a true type N burst generated by the same electron beam from three aspects of the data.
First, durations of the three branches at a given frequency increase gradually,
may due to the dispersion of the beam along its path.
Second, the flare site, as the only possible source of non-thermal electrons,
is near the western feet of large-scale closed loops.
Third, the first branch and the following two branches are
localized at different legs of the loops with opposite sense of polarization.
We also find that the sense of polarization of the radio burst is in contradiction to the O-mode
and there exists a fairly large time delay ($\sim$3-5 s) between the fundamental and harmonic components.
Possible explanations accounting for these observations are presented.
Assuming the classical plasma emission mechanism, we can infer
coronal parameters such as electron density and magnetic field near the radio source
and make diagnostics on the magnetic mirror process.

\end{abstract}


\section{Introduction}
In the solar corona, charged particles are primarily energized through two processes.
They are shock waves driven by solar eruptions and magnetic reconnections in solar flares \citep[see the reviews,][]{reames99,zharkova11}.
The accelerated non-thermal electrons 
can generate various types of drifting bursts at radio wavelengths \citep{dulk85}.
Type II and type III bursts are two most common radio emissions associated with eruptive solar activities.
Type II bursts are characterized by relatively slow-drifting narrow bands in radio dynamic spectra
and produced by energetic electrons accelerated at shocks
(see the review by Nelson \& Melrose 1985, and Chen et al. 2014; Feng et al. 2015; Kong et al., 2015, 2016, for latest studies).
On the other hand, type III bursts drift rapidly and are excited by fast electron beams (0.1-0.3 $c$)
originating from the magnetic reconnection site and escaping into the open field lines
or the large-scale closed magnetic loops \citep[see the recent review by][]{reid14}.
They are generally interpreted via the plasma emission mechanism,
with the emitting frequencies near the local plasma frequency and/or at its second harmonic \citep{ginzburg58,melrose80}.

Closed magnetic structures such as loops are ubiquitous in the lower solar corona.
If the beam of energetic electrons travels along closed loops rather than open field lines,
the frequency drift will reverse when the beam passes the loop tops and turns to the Sun.
Therefore, type J or type U bursts,
as variants of type IIIs and named because of their appearance in the radio dynamic spectra, are observed \citep{suzuki85}.
Occasionally, even a type N burst can be seen when a third branch following the U burst appears
due to further reflection of the beam via the well-known mirror effect \citep{caroubalos87}.
These fast drifting bursts are usually recorded at decimetric and metric wavelengths,
providing an important diagnostic tool for both accelerated electrons and the background media they travel through
\citep[e.g.,][]{dulk80,aschwanden92,aurass97,ning00,wang01,fernandes12,chen13,dorovskyy15}.
Early studies on type U bursts have shown that the source regions of their two branches
are usually spatially separated and associated with different parts of coronal loops or opposite field polarities \citep[e.g.,][]{suzuki78,aurass97},
while their polarization shows contradictory results.
For instance, at low frequencies ($<$200 MHz) the two branches present opposite sense of circular polarization \citep[e.g.,][]{sheridan73,suzuki78},
while at high frequencies (e.g., 327 MHz) they show the same sense of polarization
indicating a possible polarization reversal \citep[e.g.,][]{benz77,benz79}.

Type N bursts, which provide evidences for a magnetic mirror effect on the electron beams,
are more rarely observed than other type III-like bursts.
\citet{tarnstrom75} reported some radio bursts having a Y-like configuration
and may be the first observation of the reflection of electron beam at a magnetic mirror in the corona.
\citet{caroubalos87} conducted a statistical study of N-shaped radio bursts and
some events were identified as true type N bursts using the criterion of increasing duration of the three successive branches.
The increase of duration was interpreted as the dispersion of the electron beam as it travels along closed loops.
Such a spectral feature was reproduced through numerical simulations of the dynamics of electron beams in a magnetic trap \citep{hillaris88}.
Later, \citet{aurass94} and \citet{aurass96} showed that the source position of the third branch of a type N burst
is situated close to that of the second branch using Nan\c{c}ay Radioheliograph \citep[NRH;][]{kerdraon97} as expected.
However, no polarization observation of type N bursts was reported.

Earlier studies suffered from a lack of simultaneous imaging data in
the radio wavelength for the emission sources and the extreme-ultraviolet (EUV) passbands for the emitting structures.
The simultaneous availability of these two datasets is very useful to
further examine the origin of the radio bursts and
explore the potential diagnostic value on coronal parameters \citep[see, e.g.,][]{zimovets12,kumar13,tun13,bain14,chen14,feng15}.
In this paper, we report a type N burst that is well observed by
both the Atmospheric Imaging Assembly onboard the Solar Dynamics Observatory \citep[SDO/AIA;][]{lemen12}
with high-quality imaging data of the coronal structures and the relevant eruptive process
and the NRH with multi-frequency radio imaging data on the radio sources.
Such kind of combined analysis on a type N burst is conducted for the first time,
providing valuable information on the dynamic evolution and polarization change of the radio sources,
their relation with coronal structures, as well as further diagnostics on the underlying magnetic mirror process.
In Section 2, we present the observational data and major results.
In Section 3, we discuss the polarization of the radio burst and the time delay of its harmonic components.
Our conclusions are given in Section 4.

\section{Observations}
\subsection{Radio Dynamic Spectrum}

The solar radio burst of interest was recorded by several radio spectrometers.
Figures 1(a)-(b) show the radio dynamic spectra during the period 10:41:00 UT to 10:41:45 UT on 2014 May 6.
Figure 1(a) presents the data from GAURI in the range 50-250 MHz and
(b) is a composite of data from GLASGOW (50-80 MHz), OOTY (80-144 MHz) and ORFEES (144-250 MHz).
GAURI, GLASGOW and OOTY are e-Callisto\footnote[4]{\url{http://soleil.i4ds.ch/solarradio/}}
spectrometers with a time resolution of 0.25 s,
and the time resolution of ORFEES ($>$144 MHz) is 0.1 s.
We identify that the radio burst has both fundamental (F) and harmonic (H) components with a harmonic ratio $\sim$2.
Since the high frequency part ($>$80 MHz) of F component can not be discerned due to the radio interference,
we focus on the H component in further analysis.

The radio burst consists of three successive branches and looks like the letter ``N",
with the first and third branches drifting negatively and the second drifting positively.
The backbones of individual branches,
represented by the maximum intensity at a given frequency from the ORFEES data (144-190 MHz),
are shown by black pluses in Figure 1(b).
By applying a linear regression fit using the IDL function $REGRESS$,
we get the corresponding frequency drift rates as -62.4 MHz s$^{-1}$, 12.7 MHz s$^{-1}$ and -22.2 MHz s$^{-1}$,
with correlation coefficients of the fittings being 0.957, 0.996 and 0.774, respectively.
We note that the drift rate of the first branch is much larger than that of the subsequent two branches
and the second branch has the smallest drift rate.
There are several factors that may contribute to the difference in drift rate.
For instance, the apparent drift rate can be affected by the geometrical propagation effect,
i.e., the change of the direction of the beam tracing the three branches to the observer \citep{wild59,caroubalos87}.
For a beam with a speed of 0.2 $c$ (0.3 $c$), it gives a factor of 1.5 (1.9) between the beam directed to and from the observer.
Therefore, the effect may partially explain that the drift rate of the second branch is smaller than that of the third,
but it is difficult to explain the much larger drift rate of the first branch.
Other factors that may lead to the much larger drift rate of the first branch include
the gradient of electron density at two sides of the loops being different due to the asymmetry or inclination of loops,
and the loss of energy of the electrons due to collisions with the background plasma as travelling along the loops.
Note that a similar feature of drift rate can be found for the type N bursts
shown in \citet{caroubalos87} (e.g., see their Figures 3, 4, 9 and 10).

The turning frequency of the first two branches of the type N burst,
corresponding to the loop top,
ranges $\sim$110-140 MHz ($\sim$55-70 MHz) for the H (F) component.
As for the radio spectrometers noted above, ORFEES provides the best observation of the spectrum in high frequency,
especially above 180 MHz. 
The frequency range of the mirror point is estimated to be $>$200 MHz.
According to the plasma emission mechanism, the density there is higher than 1.2$\times$10$^8$ cm$^{-3}$.
If we assume there is none or constant bulk flow in coronal loops,
we have $B_2/B_1$ = $n_2/n_1$ according to the conservation of mass and magnetic flux,
where $B_{1,2}$ and $n_{1,2}$ are the magnetic field and density at the loop top and mirror point.
Thus, we can obtain a lower limit of the magnetic mirror ratio $B_2/B_1$ accounting for the type N burst to be 2.0-3.3.

As can be seen from the radio spectrum, an intriguing feature of this radio burst
is that the F component is observed a few seconds later with respect to the H component.
The F-H time delay can be confirmed by other spectrometers
that can observe both components simultaneously, e.g.,
IZMIRAN\footnote[5]{\url{http://www.izmiran.ru/stp/lars/MoreSp.html}}, San Vito and Sagamore Hill of RSTN.
To illustrate it quantitatively, in Figure 1(c),
we plot temporal profiles of radio intensity at 150 and 145 MHz (75 and 72.5 MHz)
for the H (F) component from several spectrometers and NRH. 
For NRH at 150 MHz, the maximum brightness temperature (T$_B$) in NRH images
with a time resolution $\sim$0.25 s is shown (also see Figure 4(a)).
The data of Nan\c{c}ay Decameter Array \citep[DAM;][]{lecacheux00}
with a time resolution of 1 s is used to check the low frequency spectrum.
We find that the F-H time delay is approximately 3-5 s as measured from the intensity peaks of the first branch.
This anomalous delay may arise from various propagation effects and will be discussed later in Section 3.

As mentioned earlier, the criterion of type N bursts used in \citet{caroubalos87}
is the duration of the three branches regularly increasing with time,
interpreted as the dispersion of the electron beam along its path.
For this event, as shown in Figure 1(c), we can also see increasing duration at a given frequency for the three branches.
Their half-power durations at 150 MHz are approximately 1.8 s, 4.3 s, and 12.6 s, respectively.
Thus, at this point, it agrees with the criterion of a true type N burst by \citet{caroubalos87}.
In addition, the temporal spacing of the first two branches at 150 MHz measured from the intensity peaks is $\sim$6.7 s,
and that of the last two branches is $\sim$8.8 s.
Considering the distance between active regions at the footpoints of associated closed loops (as shown below) is $\sim$0.3-0.4 $R_\odot$,
if we assume the geometry of the loops above the radio sources is semi-circular with a radius of $\sim$0.15-0.2 $R_\odot$,
we can infer the average speed of the electron beam to be $\sim$0.17-0.22 c.

\subsection{Solar Flare}

About 4 minutes before the radio burst, a compact flare
occurred in NOAA active region (AR) 12055 ($\sim$ N10E60).
Impulsive brightening is observed in all AIA EUV channels.
Figures 2(a)-(c) show AIA images in 131 \AA, 94 \AA\ and 171 \AA\ at the onset of the radio burst.
Figure 2(d) shows the magnetogram from the Helioseismic and Magnetic Imager \citep[HMI;][]{schou12} onboard the SDO
and the red contour represents 20\% of AIA 131 \AA\ intensity maximum.
We see that the major polarity of magnetic field at the flare site is negative.
Figure 2(e) displays profiles of the GOES soft X-ray (SXR) flux in 1-8 \AA\
and normalized AIA intensities at six EUV wavelengths.
The flare of C1.2 class appears as a peak in the SXR profile
during the decay phase of a M1.8 flare that started at 08:41 UT near the western limb (S17W82).
It shows that the temporal profiles of flare intensity at EUV wavelengths are generally consistent with that of the SXR.
The radio burst is generated in the impulsive phase of the C1.2 flare.

The AIA 171 \AA\ image with a larger field of view (FOV) is shown in Figure 2(f),
from which we can see some large-scale closed loops
interconnect AR 12055 and other two ARs (12056 and 12057) near/behind the eastern solar limb.
The flare occurred near the western feet of the loops in AR 12055.
Magnetic field lines obtained from the potential-field
source-surface (PFSS) extrapolation model \citep{schatten69,schrijver03} are displayed in Figure 2(g).
The large-scale magnetic field configuration above the ARs is
characterized by a group of closed loops, corresponding to the streamers as observed from coronagraphs.
In addition, according to the data of the EUV Imager \citep[EUVI;][]{wuelser04}
onboard STEREO-B ($\sim$165$^{\circ}$ behind the Earth) from another perspective,
no other flares are discernable from the three ARs during this event.
This indicates that the flare occurring in AR 12055
is the most likely source of energetic electrons for the radio burst.

\subsection{NRH Radio Imaging Data}

The radio burst was imaged by the NRH at two frequencies, i.e., 150 and 173 MHz.
Figure 3(a) shows radio source positions at the intensity peaks of each branch observed at both frequencies.
Figures 3(b)-(c) show source positions during the period of half-power intensity of each branch at 150 and 173 MHz, respectively.
The colored contours in each panel represent 90\% of the corresponding intensity maximum.
Closed field lines connecting the flare site and the radio sources derived from the PFSS model
are displayed to describe the possible path along which the electron beam travels.
We see that the first branch situates on the western side of the closed loops right above the flare,
while the following two branches appear to be close to each other and both localized on the opposite side.
This supports that the third branch of emission is a continuation of the first two branches
and excited by the same beam of electrons that are reflected via a magnetic mirror effect in coronal loops.
Taking into account the projection effect, the source heights at both frequencies are $\sim$1.4 $R_\odot$.

Figure 4(a) shows the maximum T$_B$ of the radio source in NRH images
 as a function of time at both 150 and 173 MHz.
We see that the T$_B$ is similar to that of normal type IIIs
which typically ranges from 10$^{6}$ to 10$^{12}$ K \citep{suzuki85},
and the peak intensities of individual branches are all $>$10$^{9}$ K.
At every moment, we estimate the total brightness and the circularly polarized flux, i.e., Stokes I and Stokes V,
by an integration over the contour of 50\% of the corresponding maximum T$_B$ in NRH images.
The temporal profiles of Stokes I, Stokes V and the degree of circular polarization
obtained from the ratio of Stokes V/I are presented in Figures 4(b)-(d).
There exists \textbf{an} inversion of polarization from left-handed to right-handed for the first two branches,
and the third branch remains right-handed polarization.
In addition, the degree of circular polarization is very low ($\sim$0.03-0.07).
Note that since the values of the polarization are
only marginally larger than or at the same level of the limit of the NRH measurements \citep{mercier90,kerdraon97},
the discussion on the polarization as shown below could be speculative.
Based on above observations, we see that the first branch and the subsequent two branches
are located on opposite sides of the large-scale loops,
with negative field at the western feet and positive field at the eastern feet.
Therefore, the line of sight component of magnetic field (B$_{LOS}$) in their source regions should be in opposite directions
and they are expected to be oppositely polarized.
Thus, the polarization measurement for a type N burst, first of its kind,
is consistent with the general expectation of the burst.

\section{Discussion} 
\subsection{Polarization of the H Component}
According to the plasma emission theory of type IIIs,
in the presence of magnetic field, there are two emission modes
with opposite senses of rotation of the electric field vector, i.e., the O-mode and the X-mode.
It favors O-mode emission provided that the Langmuir waves are strongly collimated along the magnetic field \citep{melrose72,melrose78,melrose80etal}.
The H radiations are generally accepted to be in the sense of the O-mode \citep[e.g.,][]{dulk80,suzuki85}.
Therefore, the polarization is expected to be left-handed (right-handed) in positive (negative) magnetic field.
Obviously, as shown in Figure 4, the polarization of the type N burst is in contradiction to this expectation.
One possibility is that the H radiation of the type N burst is polarized in the X-mode
which may occur under certain conditions \citep[e.g.,][]{willes97}.
On the other hand, if the radio burst indeed corresponds to the O-mode,
it can be interpreted by a polarization reversal along the ray path.

In the theory of mode coupling, the sense of polarization changes
when the ray crosses a so-called quasi-transverse (QT) region if the mode coupling is weak \citep{cohen60,suzuki85}.
A QT region refers to where the magnetic field is quasi-perpendicular to the ray path
and the B$_{LOS}$ changes its direction (see Figure 4(e)).
The theory has been used to interpret the same sense of polarization
observed in type U bursts and other radio emissions from bipolar regions \citep[e.g.,][]{melrose73,benz77,benz79,white92,gopalswamy94}.
The weak coupling condition is $f^4 \ll f_t^4$, where $f_t$ is the transition frequency, $f_t \propto (n_e B^3 L)^{1/4}$,
 $n_e$ is the coronal density and $L$ is the characteristic length of the QT region.
By using the same coronal density and magnetic field models and the scale length as \citet{benz77},
we can estimate the maximum height below which the weak coupling condition is satisfied.
For radio emissions at 150 and 173 MHz, we find the maximum height to be $\sim$2-2.5 $R_\odot$.
It is not easy to determine the existence of a QT region along the ray path
since we do not have direct observation of the magnetic field in the corona.
Based on observations from the EUV images and white-light coronagraphs,
the radio sources are embedded in a group of large-scale loops near the eastern limb,
as shown by the schematic in Figure 4(e).
We suggest that the radio emission is likely to pass through an overlying region in the corona
where the magnetic field is quasi-perpendicular to the ray path.
In consequence, the original right-handed (left-handed) polarization changes to left-handed (right-handed),
 while the sense of O mode retains.

As shown in Figure 4(d), the radio burst is only weakly polarized, with the degree of circular polarization $\sim$0.03-0.07.
The weak polarization is consistent with previous observations of type U bursts or the H components of type IIIs
\citep[e.g.,][]{benz77,benz79,dulk80,mercier90,aurass97}.
\citet{dulk80} showed that the average degree of polarization is 0.11 for H components of type IIIs
and 0.06 for structureless type IIIs (with no observable F-H feature),
 and the polarization decreases from the disk center to the limb.
 \citet{aurass97} studied a sample of 23 type U bursts
 and found that the degree of polarization is generally low ($<$0.1) and difficult to measure.
Considering that our event occurred near the limb (flare site $\sim$ E60), weak polarization is expected.
The degree of polarization for the H component can be calculated with the equation $p = a(\theta) f_B/f_p$,
where $a$ is a factor depending on the angle $\theta$ between the magnetic field and the viewing direction,
$f_B$ and $f_p$ are gyro-cyclotron and plasma frequencies (see \citet{suzuki85}, and references therein).
Therefore, we can use it to estimate the magnetic field strength near the radio source \citep[e.g.,][]{dulk80,reiner07,ramesh10,sasikumar13}.
For our case, since $\theta$ is likely $>$60$^{\circ}$, we can take $a \sim$0.5-1
and thus obtain the magnetic field $B \sim$0.8-3.8 Gs.
The value generally agrees with that deduced from the empirical model of \citet{dulk78}, which gives $B \sim$2.0 Gs at 1.4 $R_\odot$.
From the PFSS extrapolation, we also find the magnetic field is approximately 1-2 Gs at 1.4 $R_\odot$ above the ARs.

\subsection{Time delay of F-H components}
As noted above, the F component of the type N burst is delayed by $\sim$3-5 s relative to the H component.
Similar time delay of a few seconds has been observed for type III and type U bursts \citep[e.g.,][]{stewart74,suzuki85}.
However, to date, there is no satisfactory quantitative explanation,
since many effects may be involved during the propagation of radio emissions.
The first effect is the relatively lower group velocity of the F radiation near the radio source
as it is emitted near the plasma frequency.
The time delay caused by this effect is usually estimated to be $\lesssim$1 s at meter wavelength \citep[e.g.,][]{robinson98},
thus inadequate to account for the observation.
Refraction and scattering of radio emission can lead to time delay as well \citep[e.g.,][]{leblanc73,riddle74}.
In the numerical simulation of type IIIs, \citet{li08} showed that the H emission arrives earlier than the F,
but the time difference is only 0.7 s for 200 MHz (H).
Note that the group speed difference also contributes in the simulation.
Another possible effect is the wave ducting in low-density flux tubes \citep[e.g.,][]{stewart74,duncan79}.
In this scenario, both components are excited within and ducted along the flux tube,
 while the H component escapes at a lower height and thus appears earlier in the radio spectrum.

Recently, \citet{dorovskyy15} found a time delay as long as 7 s for a decametric type U burst harmonic pair.
They suggested that the time delay arose from the different group velocities of F-H components,
and the radio source was emitted within large-scale loop structures
with heights of 4-6 $R_\odot$ and containing plasmas of high density and temperature.
Such high loop structures were assumed to be created by a coronal mass ejection (CME).
Note that, however, radio bursts in this frequency range ($>$25 MHz) usually originate below $\sim$2 $R_\odot$.
For our event, we can not find such high-lying structures as proposed by \citet{dorovskyy15} to account for the time delay.
We suggest that the anomalous F-H time delay of the radio burst
can not be explained by a simple theory and may be the consequence of combined propagation effects.

\section{Conclusions}

Type N bursts reveal a magnetic mirror effect on the beam of energetic electrons travelling along coronal loops.
In this paper, we present one such event well observed by both the AIA at the EUV wavelength and the NRH at metric wavelength.
This kind of combining data analysis is done for the first time.
We provide three aspects of evidence to support the picture of a type N burst.
First, durations of the three successive branches at a given frequency increase gradually.
Second, the flare associated with the burst occurred near the western feet of a group of large-scale closed loops.
Third, the first branch and the subsequent two branches are
localized on different sides of the loops with opposite sense of polarization.

The apparent drift rates of the three branches of the type N burst are found to be quite different.
The smaller drift rate of the second branch than that of the third branch
could be partially explained by the geometrical propagation effect,
but it is difficult to interpret the much larger drift rate of the first branch.
It suggests that other factors should also have effect on the different drift rates of the three branches.
In addition, from the dynamic spectrum, we find that the magnetic mirror accounting for the type N burst
is characterized by a lower-limit mirror ratio of $\sim$2.0-3.3.

Generally, it is believed that the H radiations of type IIIs or type III-like bursts belong to the O-mode.
By examining the field polarities at the feet of the corresponding loops,
we find that the sense of polarization of the type N burst, reported for the first time,
conflicts with the expectation of O-mode emission.
We suggest that the radio emissions located on both sides of closed loops
possibly pass through a QT region below $\sim$2.5 $R_\odot$ where the mode coupling is weak.
As a result, their original handedness of polarization is reversed,
 but they remain oppositely polarized.
From the polarization measurement of the H component,
we can also estimate the magnetic field strength near the radio source,
which is $\sim$0.8-3.8 Gs at $\sim$1.4 $R_\odot$.

We also find that the F component of the radio burst is observed $\sim$3-5 s later than the H component.
At present, there is no satisfactory explanation for the F-H time delay,
although it has been noted for a long time.
Several propagation effects that may play a role have been proposed,
including the group velocity difference of F-H radiations near the radio source,
refraction and scattering of radio emissions, and the wave ducting effect in low-density flux tubes.
However, none of those propagation effects can easily predict a time delay of several seconds.
We suggest that the anomalous F-H time delay of the radio burst
may be the contribution of different propagation effects.

The type N burst reported here is perhaps the best observed of its kind.
Yet the comprehensive observations have served to starkly reveal gaps in our understanding of
the beam-driven plasma emission in the solar corona.
To better interpret the observational characteristics of the type N burst,
numerical efforts on the dynamics of electron beams and the coherent plasma emission are needed,
in addition to making more multi-wavelength observations of this kind of events.

\acknowledgements

We thank the teams of e-Callisto, ORFEES, NRH, DAM, IZMIRAN, RSTN, SDO/AIA and SDO/HMI for making their data available to us.
This work was supported by grants NSBRSF 2012CB825601, NNSFC 11503014, 41274175, 41331068, U1431103,
and Natural Science Foundation of Shandong Province ZR2014DQ001.
Gang Li's work at UAHuntsville was supported by NSF grants ATM-0847719 and AGS1135432.

\begin{figure}
\includegraphics[width=0.85\textwidth,clip,trim=0cm 0cm 0cm 1.6cm]{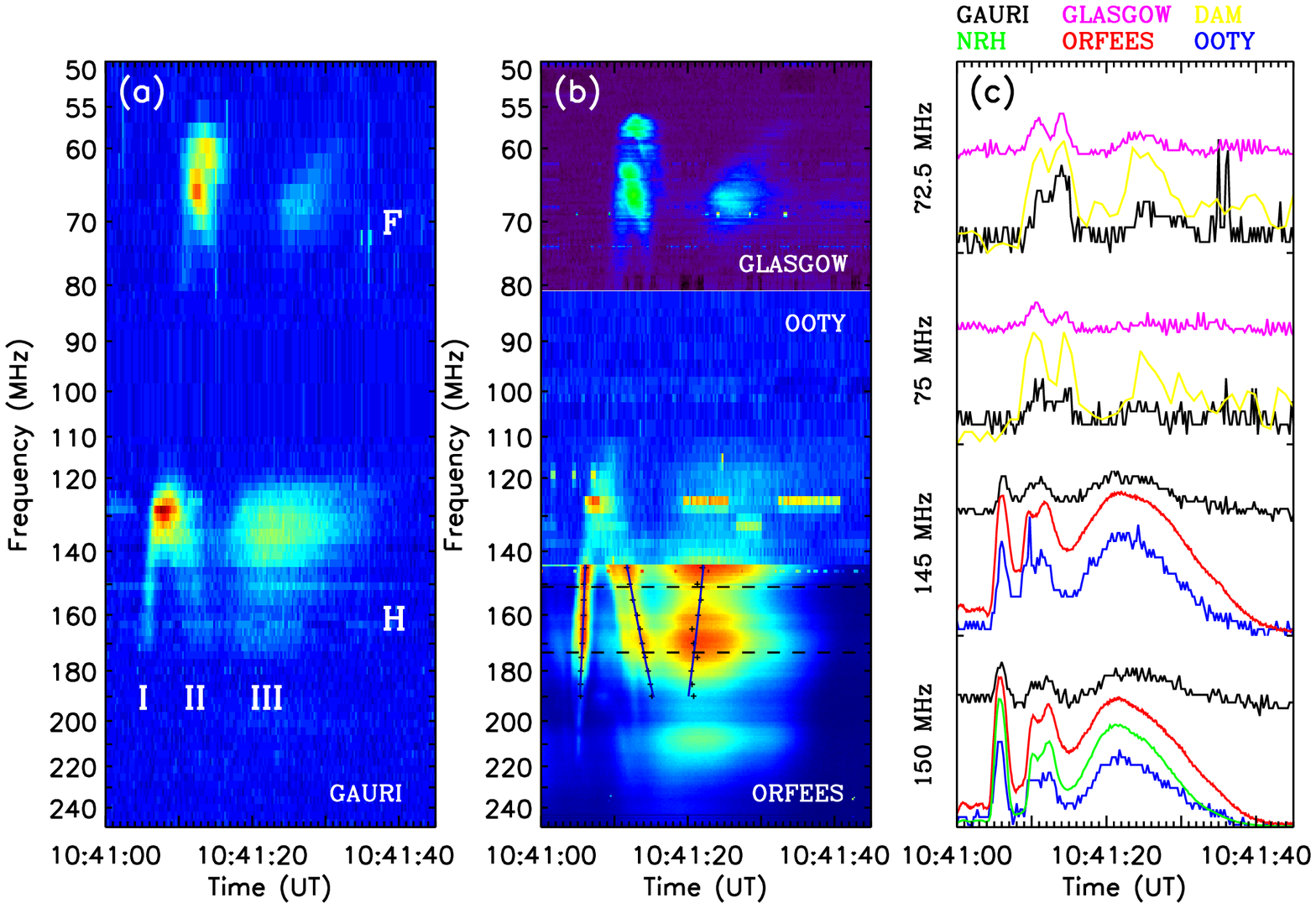}
\caption{(a) Radio dynamic spectrum recorded by GAURI in the range 50-250 MHz.
``I", ``II" and ``III" denote the three branches of the radio burst,
and ``F" and ``H" denote the fundamental and harmonic components.
(b) A composite spectrum from GLASGOW (50-80 MHz), OOTY (80-144 MHz) and ORFEES (144-250 MHz).
The black pluses show the profiles of individual branches by maximum intensity at a given frequency using the ORFEES data
and the blue lines are obtained by linearly fitting the data points.
Dashed lines indicate the two NRH imaging frequencies (150 and 173 MHz).
(c) Temporal profiles of radio intensity at four frequencies from several instruments are plotted in arbitrary unit.
Note that the spectra of GLASGOW and OOTY have been shifted by 1.5 s and 103 s respectively
to keep the spectral data consistent.
}\label{Fig1}
\end{figure}

\begin{figure}
\includegraphics[width=0.95\textwidth,clip,trim=0cm 2cm 0cm 3cm]{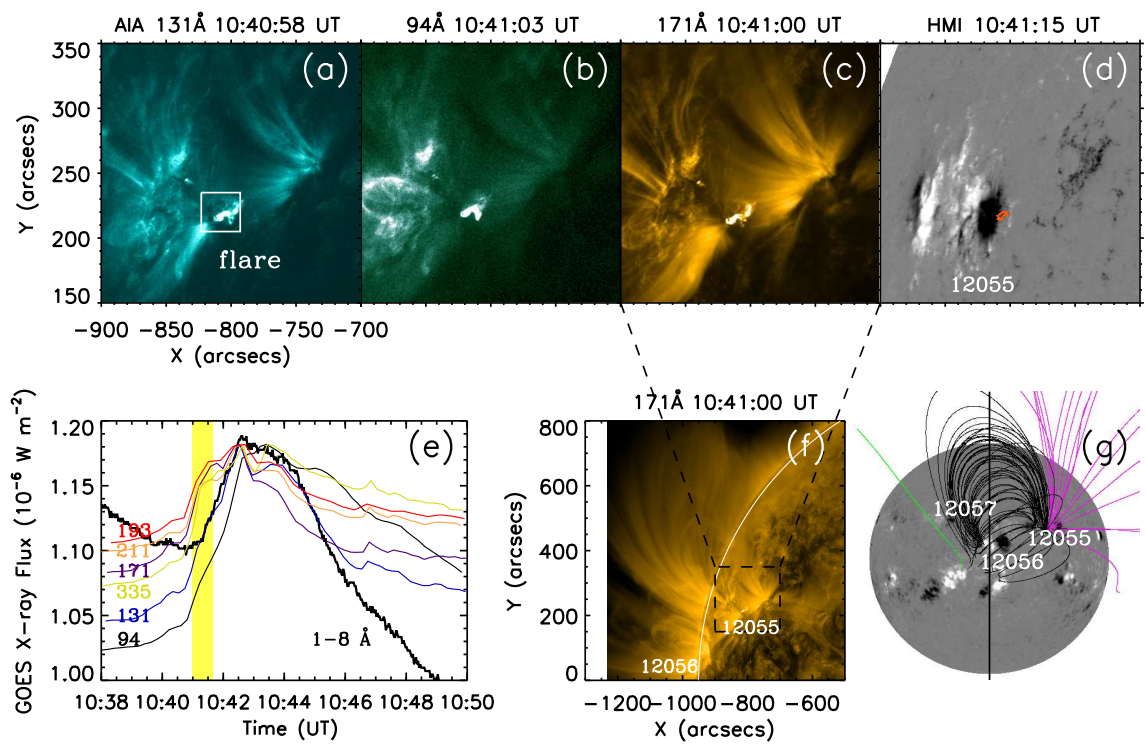}
\caption{(a)-(d) AIA 131 \AA, 94 \AA\ and 171 \AA\ images and HMI magnetogram.
 Red contour in panel (d) represents 20\% of the AIA 131 \AA\ intensity maximum.
(e) Temporal profiles of GOES soft X-Ray flux in 1-8 \AA\ (thick black line) and
normalized AIA intensities integrated over the white box marked in panel (a) at six wavelengths (lines in different colors).
The shaded region indicates the period of radio emission.
(f) AIA 171 \AA\ image with a larger field of view.
(g) Magnetic field configuration obtained from the PFSS model and rotated 90$^{\circ}$ to the west.
The black lines represent closed field lines and
the vertical black line denotes the position of solar limb as viewed from the Earth.
}\label{Fig2}
\end{figure}

\begin{figure}
\includegraphics[width=0.95\textwidth,clip,trim=0cm 0cm 0cm 0cm]{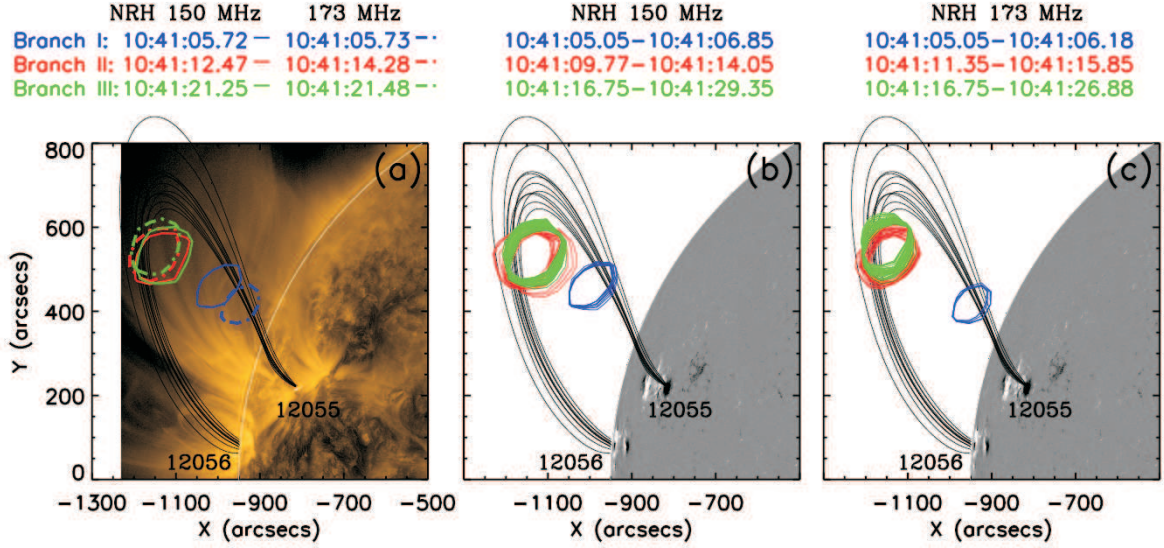}
\caption{Radio source positions imaged by the NRH 
at the intensity peaks of each branch at both 150 (solid) and 173 (dash-dotted) MHz (a),
during the period of half-power intensity of each branch at 150 MHz (b) and 173 MHz (c).
The contours in each panel represent 90\% of the corresponding intensity maximum
and the three branches are plotted in blue, red and green, respectively.
Closed field lines connecting the flare site and radio sources
 obtained from the PFSS model are superimposed on the AIA 171 \AA\ image in panel (a) and HMI magnetogram in panels (b)-(c).
}\label{Fig3}
\end{figure}

\begin{figure}
\includegraphics[width=0.95\textwidth,clip,trim=0cm 0cm 0cm 0cm]{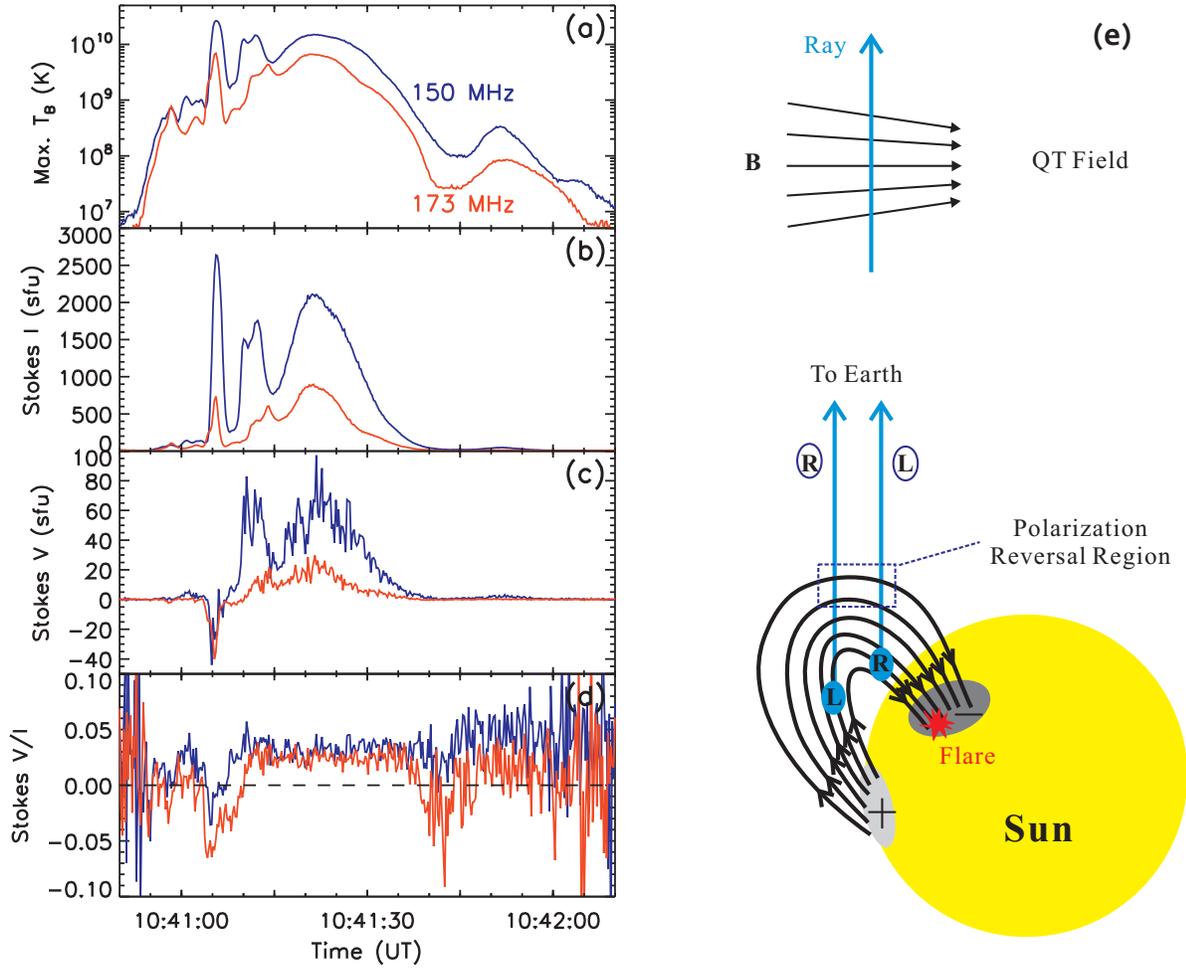}
\caption{Left: Temporal profiles of the maximum T$_B$ (a), Stokes I (b), Stokes V (c),
and the degree of circular polarization Stokes V/I (d) at 150 (blue) and 173 (red) MHz recorded by the NRH.
Right: Schematics of the QT field region and the polarization reversal when
radio emissions pass through a QT region and the weak mode coupling is satisfied.
}\label{Fig4}
\end{figure}

\end{document}